\documentclass[conference]{IEEEtran}
\usepackage{epsfig}
\usepackage{graphicx}
\usepackage{color}
\usepackage{subfigure}

\usepackage{amssymb}
\usepackage{makeidx}
\usepackage{amsmath}
\usepackage{amsthm}
\usepackage{graphicx}
\usepackage{plain}
\usepackage{epsf}
\usepackage{epsfig}
\usepackage{psfig}
\usepackage{ccaption}
\usepackage{array}
\usepackage{tabularx}
\usepackage{multirow}
\usepackage{epsfig}
\usepackage{cite}
\usepackage{procedure,algorithm,algorithmic}

\begin{document}
%
\title{Time Stamp Attack in Smart Grid: Physical Mechanism and Damage Analysis}

\author{\IEEEauthorblockN{Shuping Gong,
Zhenghao Zhang,
Husheng Li}

\IEEEauthorblockA{Department of Electrical Engineering and Computer Science\\
University of Tennessee,
Knoxville, TN 37996\\ Email: sgong1@utk.edu; zzhang26@utk.edu; husheng@eecs.utk.edu}
\and
\IEEEauthorblockN{Aleksandar D. Dimitrovski}

\IEEEauthorblockA{Energy and Transportation Sciences Division\\
Oak Ridge National Lab,
Oak Ridge, TN 37831\\ Email: dimitrovskia@ornl.gov}}

\maketitle

\begin{abstract}
Many operations in power grids, such as fault detection and event location estimation, depend on precise timing information. In this paper, a novel time stamp attack (TSA) is proposed to attack the timing information in smart grid. Since many applications in smart grid utilize synchronous measurements and most of the measurement devices are equipped with global positioning system (GPS) for precise timing, it is highly probable to attack the measurement system by spoofing the GPS. The effectiveness of TSA is demonstrated for three applications of phasor measurement unit (PMU) in smart grid, namely transmission line fault detection, voltage stability monitoring and event locationing.
\end{abstract}

\section{Introduction}
Smart grid \cite{Grid_future} has been considered as an emerging technology profoundly changing the modern power grids. To maintain the reliability of power systems, wide area monitoring systems (WAMSs) \cite{FNET_SGTRAN} are exploited to obtain the real-time system status, which is essential for the maintenance and control of power systems. The security of WAMSs is one of the prime issues in the smart grid technology, since the power grid will make operation decisions depending on these measurements from the WAMS. Errors of measurements will cause wrong operations that may lead to serious damages such as instability or even blackout.

Most studies on security issues in smart grid have been focused on how to protect the data integrity of the measurements. Accordingly, the attack on the measurements is named the false data injection attack (FIA) \cite{LiuYao_FDI}\cite{Poor_SGCOM_11}. Different from FIA, which requires hacking into the computer system of power grid, in this paper, we identify a potential type of attack to the WAMS, coined {\it Time Stamp Attack (TSA)}, which occurs in the physical layer.

\begin{figure}[]
  \centering
  \includegraphics[scale=0.4]{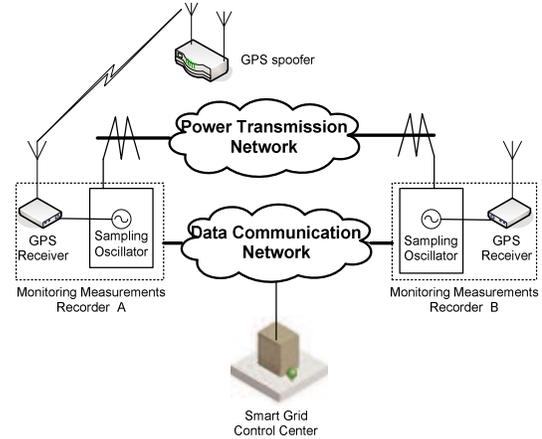}
  \caption{Illustration of synchronized smart grid monitoring with GPS spoofer}\label{Fig:SynMeasure}
\end{figure}

It is well known that the measurements in the WAMS need to be time synchronized \cite{Daggle_synch}\cite{FNET_SGTRAN}, which is often achieved by using the global positioning system (GPS).
As illustrated in Fig. \ref{Fig:SynMeasure}, with a GPS signal receiver, monitoring measurement recorders (MMRs) trigger their measurements by the GPS time signal. After the measurements are recorded,
the time values are attached to the measurements,
which is similar to posting a stamp to the measurements (thus called time stamp).
Through the communication infrastructure, the measurements with time stamps are conveyed to the control center, based on which the control center can align the collected measurements for analyzing the system state and then take future actions.

Although providing important information, the time stamps are vulnerable to attacks, since they can be modified by inducing a forged GPS signal \cite{Spoofer}.
Moreover, it is difficult for a common civil GPS receiver to detect a spoofing GPS signal.
The attack can be implemented successfully with a high probability,
since the attacker even does not need to hack into the monitoring system.
Although there is some data processing procedure to handle the measurements, most current processing schemes only consider the measurement data error caused by noise and apply a simple smoothing filtering scheme.
Consequently, TSA can easily bypass the data processing procedure.

In this paper, the impact of TSA will be evaluated for three applications of PMU, namely transmission line fault detection/locationing, voltage stability monitoring and event locationing. Simulation results will demonstrate that TSA can effectively deteriorate the performance of these applications and may even result in false operation of power system.

The remainder of this paper is organized as follows.
Section \ref{sec: Attack model} provides the GPS spoofing attack model from the aspect of signal processing.
Section \ref{sec:app} briefly introduces the backgrounds for applications of PMUs including
the transmission line fault detection/locationing algorithm,  the voltage monitoring algorithm and the event locationing.
Simulations for demonstrating the impacts of TSA are presented in Section \ref{sec:sim}. Conclusions and future work are provided in Section \ref{sec: conclusion}.

\section{GPS Signal and Attack Model}\label{sec: Attack model}
In this section, we briefly introduce the GPS signal reception processing. Then we propose the attack model for GPS spoofing and TSA.

\subsection{GPS Signal Reception}
The precise timing information from GPS signals includes two parts. One is embedded in the navigation messages demodulated from the received GPS signals, which has the precision of a second; the other part is the precise signal propagation time from the GPS satellite to the receiver, which has the precision of a millisecond for civil users.

The system-wide synchronization time reference is referred to the coordinated universal time (UTC) $t_{UTC}$ disseminated by GPS, which is given by
\begin{equation}\label{Eq. UTC}
    t_{UTC} = t_{rcv}-t_{p}-\Delta t_{UTC},
\end{equation}
where $t_{rcv}$ and $t_{p}$ represent the receiver clock time and propagation time for the GPS signal, respectively,
and $\Delta t_{UTC}$ denotes the time corrections provided by the GPS ground controllers.
To obtain the navigation message, we need to demodulate the GPS signal.

The received standard positioning service (SPS) GPS signal $r(t)$ is given by
\begin{equation}\label{Eq. GPS_signal}
    r(t) = \sum_{k=1}^{32}H_k(2P_c)^{1\over 2}(C_k(t)\oplus D_k(t))cos2\pi (f_{L1}+\Delta f_k)t+n(t),
\end{equation}
where $H_k$ and $P_c$ are the channel matrix for the $k$-th satellite and the signal power, respectively,
$C_k(t)$ and $D_k(t)$ are the spread spectrum sequence (C/A code) and the navigation message data from the $k$-th satellite, respectively,
$f_{L1}$ and $\Delta f_k$ are the carrier frequency for civil GPS signal and doppler frequency shift for the $k$-th satellite, respectively, and $n(t)$ denotes the noise.
The received signal processing includes two major steps, namely acquisition and tracking.
>From (\ref{Eq. GPS_signal}), we can observe that the key processing for acquisition is to search for the code phase of the receive
C/A code and doppler frequency shift $\Delta f_k$. By multiplying the C/A code of identical code phase with the carrier of the same frequency as the received GPS signal, the navigation message can be demodulated coherently \cite{Book_GPSreceiver}.

\subsection{Attack Model}
\begin{figure}[]
\vspace{0pt}
\subfigure[No attack\label{Corr_ture}]{
\begin{minipage}[tpb]{2cm}
\centering
\includegraphics[scale=0.3]{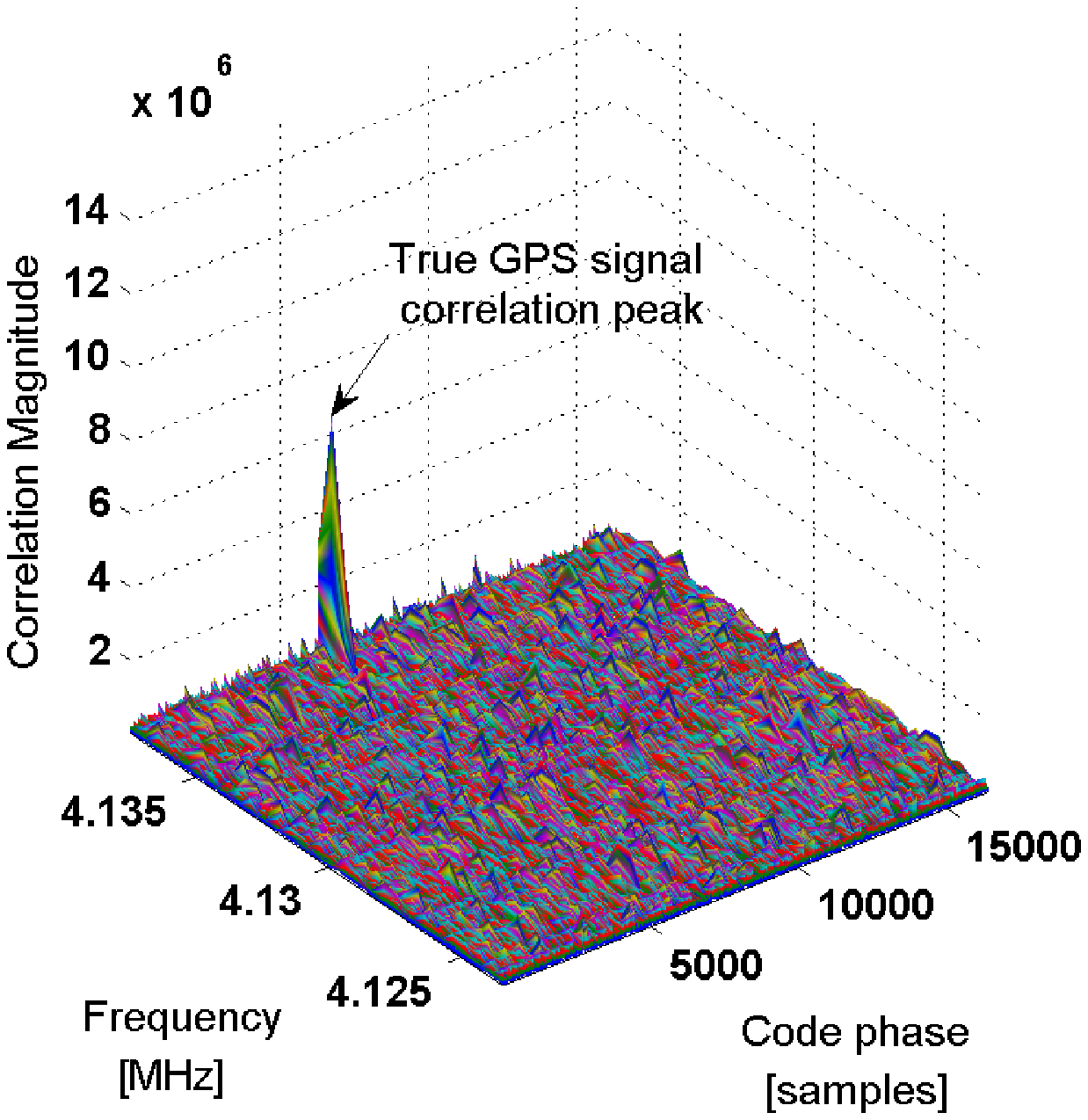}
\end{minipage}}%
\hfill \vspace{0pt} \subfigure[Under spoofing attck\label{Corr_attack}]{
\begin{minipage}[tpb]{4cm}
\centering
\includegraphics[scale=0.3]{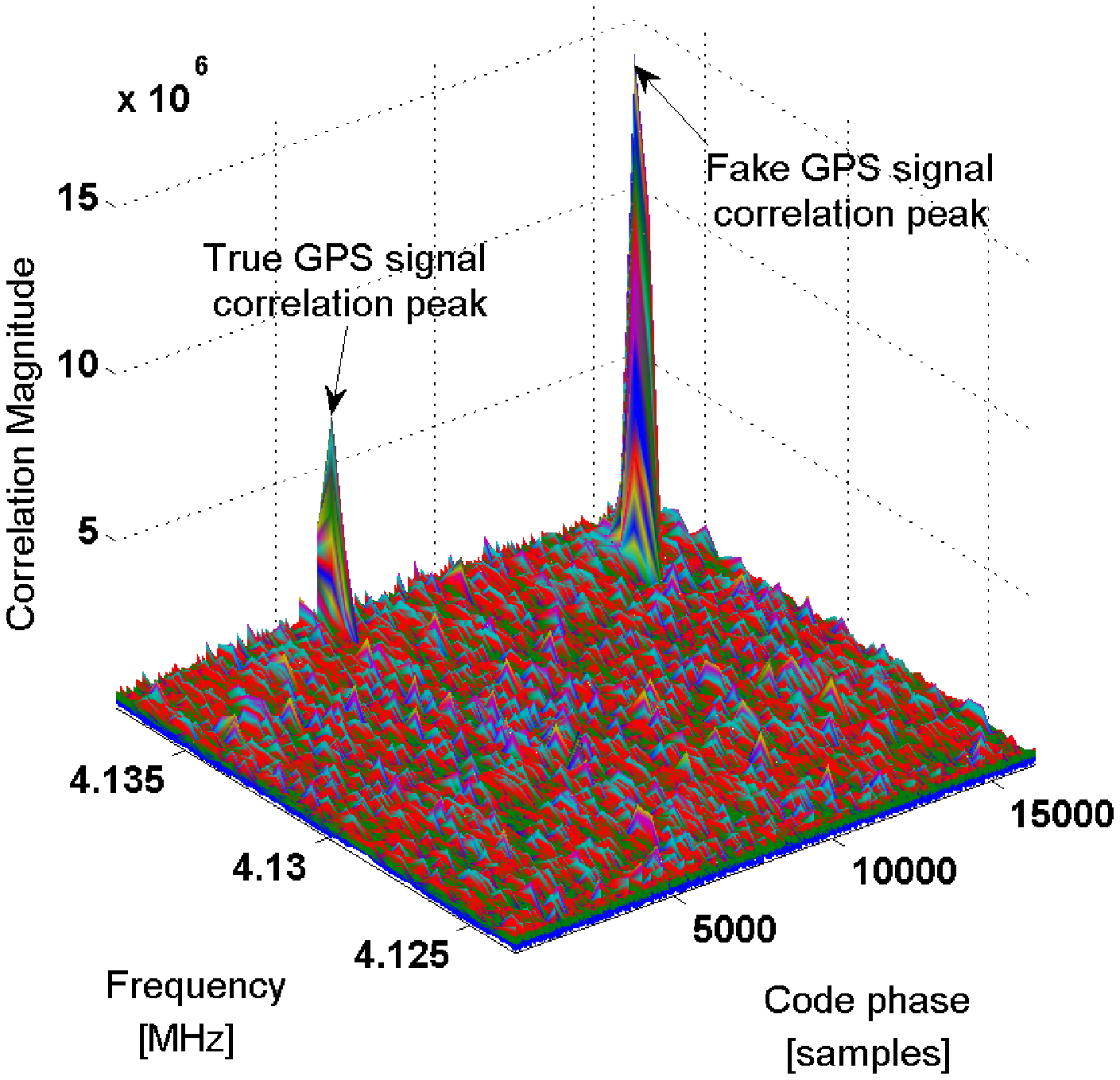}
\end{minipage}}
\caption{Comparison of correlation peak under normal and spoofing attack conditions.} \label{Fig:CorrPeak}
\end{figure}

To spoof a GPS receiver, the GPS receiver needs to be misled to acquire the fake GPS signal instead of the true one.
The acquisition is implemented by searching for the highest correlation peak in the code phase-carrier frequency two dimensional
space. Intuitively, the signal with a higher signal-to-noise-ratio (SNR) will have a higher correlation peak, which is illustrated in Fig. \ref{Fig:CorrPeak}.

Therefore, there exists a two-step spoofing strategy. In the first step, the spoofer launches a certain interference and causes the GPS receiver to lose track. Then, it sends spoofing GPS signal when the GPS receiver carries out the acquisition processing. Consequently, the GPS receiver will track the fake GPS signal due to its higher correlation peak, since the fake GPS signal has a higher SNR.

\section{Introduction to Applications of PMU}\label{sec:app}
In this section, we provide a brief introduction to the applications of PMUs requiring precise timing information, which include
the transmission line fault detection/locationing algorithm,  the voltage monitoring algorithm and the event locationing.

\subsection{Transmission Line Fault Detection and Locationing}\label{subsec:fault}
An algorithm for transmission line fault detection/locationing can quickly detect the fault and estimate the fault location. Many studies have suggested to utilize the measurements at both ends of the transmission line to improve the locationing accuracy \cite{Novosel1996, Jiang2000, Liao2007}. Here, we briefly review the fault detection and locationing method proposed in \cite{Jiang2000} which utilizes the measurements at both ends and thus requires time synchronization.  As \cite{Jiang2000} focuses on long transmission lines, we can extend it to short and medium transmission line, which is omitted in this paper due to the limited space.



\begin{figure}[htpb]
  \centering
  \includegraphics[scale=0.55]{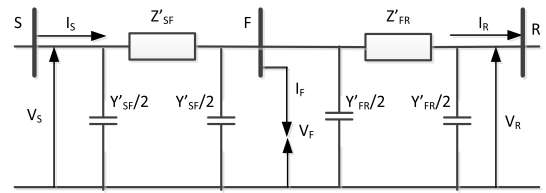}
  \caption{Model for long transmission line with fault}\label{fig:m_fault_l}
\end{figure}

Fig. \ref{fig:m_fault_l} shows the long transmission line model \cite{Abboud1} with fault \cite{Jiang2000}. In this model $V_R^\prime$ and $V_S^\prime$ are the voltages at the receiving and sending ends with unit $V$; $I_R^\prime$ and $I_S^\prime$ are the currents at both ends of the line with unit $A$. Suppose that the total length of transmission line is $L$ miles or kilometers. The length from the fault to the receiving end is $DL$ miles or kilometers, in which $D$ is the fault location index. $Z_{SF}^\prime$ and $Z_{FR}^{\prime}$ are the impedances of each transmission line segment. $Y_{SF}^\prime$ and $Y_{FR}^{\prime}$ are the admittances of each transmission line segment.

When a fault occurs, the voltage $V_F$ at the fault location can be calculated from the measurements, $V_{SF}^\prime$ and $I_{SF}^\prime$, at the sending side, or measurements $V_{FR}^\prime$ and $I_{FR}^\prime$, at the receiving side. The computation results of $V_F$ from both sides should be equal to each other. Based on this observation, the fault locationing index can be estimated as \cite{Jiang2000}:
\begin{eqnarray}
D_e = \frac{\ln(N/M)}{2\gamma L},
\end{eqnarray}
where
\begin{eqnarray}\label{eq:comp_M_N}
N = \frac{V_R - Z_cI_R}{2} - \frac{V_S - Z_cI_S}{2}\exp(\gamma L), \\
M = \frac{V_S + Z_cI_S}{2}\exp(-\gamma L) - \frac{V_R + Z_cI_R}{2},
\end{eqnarray}
where $Z_c$ is called the characteristic impedance of the line which is equal to $Z_c = \sqrt{z_1/y_1}$. As $N$ and $M$ change suddenly because of the fault and their post-fault values are greatly larger than the pre-fault one, $N$ and $M$ can be used as the fault indicators \cite{Jiang2000}. Note that the above computation needs a perfect time synchronization of the measurements at the two ends, which exposing the system to possible TSA.


\subsection{Voltage Stability Monitoring}\label{sec:vol}


One commonly used method to evaluate the voltage stability is to use the Thevenin Equivalent Circuit to simplify the model \cite{Larsson2002}. The principle is to model the remote system as a voltage source $\bar{E}_{th}$ with impedance $\bar{Z}_{th}$, and the local load as an impedance $\bar{Z}_L$. The maximum power can be obtained when $|\bar{Z}_{th}| = |\bar{Z}_L|$.

\begin{figure}[htpb]
  \centering
  \includegraphics[scale=0.5]{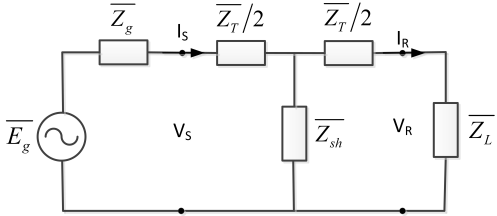}
  \caption{Transformed circuit for Power System}\label{fig:vol_eq2}
\end{figure}

With the Thevenin Equivalent Circuit, two indices for voltage stability margin can be obtained \cite{Larsson2002}.
The first index is associated with the load impedance:
\begin{eqnarray}
\text{MARGIN}_Z = 100 (1 - k_{\text{crit}}),
\end{eqnarray}
where
\begin{eqnarray}
k_{\text{crit}} = \left|\frac{\bar{Z}_{th}}{\bar{Z}_L}\right|.
\end{eqnarray}
The second index is associated with active power delivered to the  load bus (in p.u.)
\begin{eqnarray}
\text{MARGIN}_P = \left\{
\begin{array}{ll}
p_{\text{Lmax}} - P_L & \text{if} \bar{Z}_L > \bar{Z}_{th} \\
0 & \text{if} \bar{Z}_L > \bar{Z}_{th}
\end{array}
\right.
\end{eqnarray}
The indices can be used to evaluate the stability. More details can be found in \cite{Larsson2002}. Note that the above parameters need to be estimated from the synchronized measurements, thus making the algorithm vulnerable to possible TSA.

\subsection{Event Locationing}
One of the essential monitoring tasks in smart grid is to locate the disturbing event in power grid. When a significant disturbance occurs, there will be many symptoms such as voltage and frequency violations in both time and space.
The perturbation will travel throughout the grid \cite{V_event}.
Therefore, the distributed monitoring devices will capture the variance of the measurements and send these data to the monitoring system server or exchange with their neighbors. The event time and location can be deduced from the time stamp on these measurements.
After receiving the measurements from these monitoring devices, servers need to decide the hypocenter of the event, which is typically marked as the wave front arrival time \cite{Event_7}.
On aligning these measurements according to their time stamps, the event arriving time at each monitoring device can be attained. Consequently, the disturbing event location can be deduced by triangulation, which is given by (consider four MMRs)
\begin{eqnarray}\label{Eq. location}
  (x_1-x_e)^2+(y_1-y_e)^2-V_e^2(t_1-t_e)^2 &=& 0 \nonumber\\
  (x_2-x_e)^2+(y_2-y_e)^2-V_e^2(t_2-t_e)^2 &=& 0 \nonumber\\
  (x_3-x_e)^2+(y_3-y_e)^2-V_e^2(t_3-t_e)^2 &=& 0 \nonumber\\
  (x_4-x_e)^2+(y_4-y_e)^2-V_e^2(t_4-t_e)^2 &=& 0,
\end{eqnarray}
when $t_i, i=1,2,3,4$ is the disturbing event arrival time to the $i$-th MMR, $(x_i,y_i)$ and $(x_e,y_e)$ are the coordinates of the $i$-th MMR and the disturbing event location, respectively, and $V_e$ is the event propagation speed in the power grid network. Since the coordinates and the arrival time of each MMR are known, the Newton's method can be applied to solve these equations and attain the event location and time. Obviously, if the timing is incorrect, a wrong event location will be deduced from the incorrect equations.

\section{Damage of Time Stamp Attack}\label{sec:sim}

In this section, simulations have been conducted to evaluate the damage of TSA on the three applications of PMUs introduced in the previous section. Since the main impact of TSA on smart grids is the asynchronism of phase angle measurements among PMUs, we focus on evaluating the impact of the asynchronism on these applications. The phase angle errors resulted from TSA at the sending PMU and receiving PMU are denoted by $\Delta\theta_S$ and $\Delta\theta_R$, respectively. The phase angle asynchronisim between the sending PMU and receiving PMU is denoted by $\Delta\theta$ which is equal to $\Delta\theta_R - \Delta\theta_S$.

\subsection{TSA on Transmission Line Fault Detection and Locationing}

\begin{figure}
  \centering
  \includegraphics[scale=0.45]{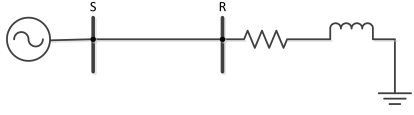}
  \caption{Simulation model for transmission line fault locationing}\label{fig:sim_f_model}
\end{figure}

The simulation model for transmission line is shown in Fig. \ref{fig:sim_f_model}. The parameters for the transmission line are the same as those in \cite{Novosel1996}. The lengths for long, medium and short transmission lines are 400 miles, 50 miles and 25 miles, respectively. The total simulation time is 10s, and the fault occurs at 5s.

\subsubsection{Short Transmission Line}
\begin{figure}
  \centering
  \includegraphics[scale=0.35]{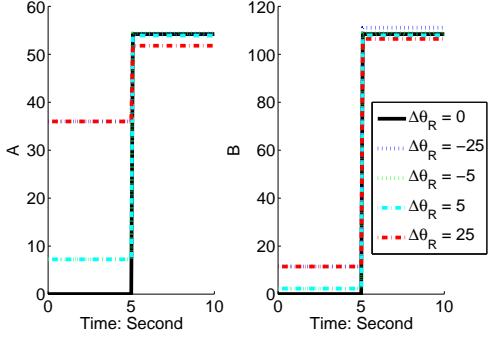}
  \caption{Fault indicators, $A$ and $B$, for short transmission line}\label{fig:sim_fs_ab}
\end{figure}

\begin{figure}
  \centering
  \includegraphics[scale=0.35]{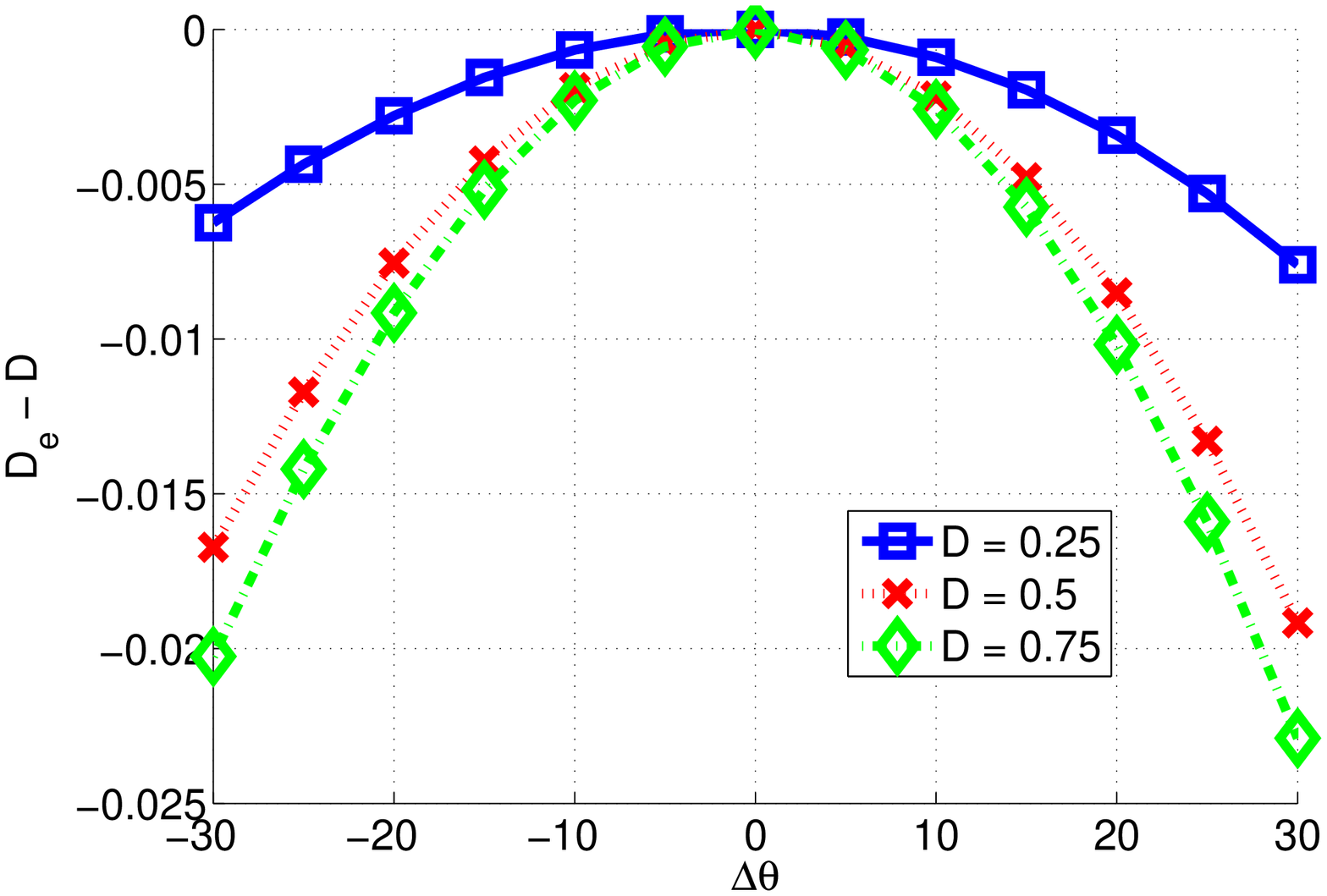}
  \caption{Performance of fault locationing for short transmission line}\label{fig:sim_fs_error}
\end{figure}

Fig. \ref{fig:sim_fs_ab} and Fig. \ref{fig:sim_fs_error} show the fault indicators, $A$ and $B$ (the computational details will be given in our journal version), and the performance of fault locationing for short transmission line with different phase angle asynchronism $\Delta\theta$. Fig. \ref{fig:sim_fs_ab} demonstrates that the gaps for fault indicators, $A$ and $B$, decrease as $|\Delta\theta|$ increases. For $A$, the gaps corresponding to $|\Delta\theta| = 0, 5, 25$ are around 55, 45, and 20, respectively. In other words, if $A$ is used as the fault indicator, the performance of fault detection will be deteriorated by TSA. As shown in Fig. \ref{fig:sim_fs_error}, the fault locationing error is very small even if $\Delta\theta$ is as large as 30. Therefore, the performance of fault locationing for short transmission lines is only negligibly affected by TSA.

\subsubsection{Medium Transmission Line}

\begin{figure}
  \centering
  \includegraphics[scale=0.35]{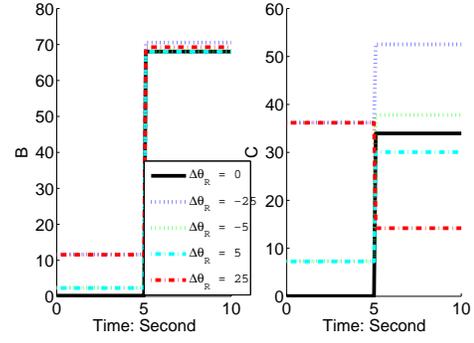}
  \caption{Fault indicators, B and C, for medium transmission line}\label{fig:sim_fm_bc}
\end{figure}

\begin{figure}
  \centering
  \includegraphics[scale=0.35]{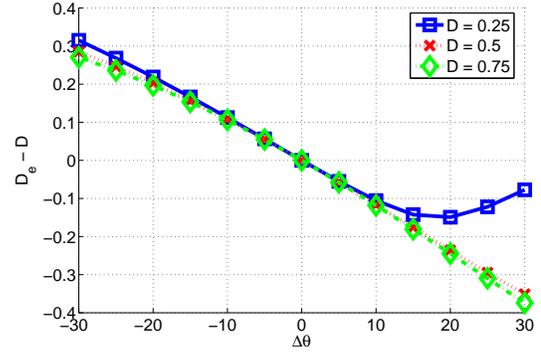}
  \caption{Performance of fault locationing for medium transmission line}\label{fig:sim_fm_error}
\end{figure}

Fig. \ref{fig:sim_fm_bc} and Fig. \ref{fig:sim_fm_error} depict the fault indicators, $B$ and $C$ (the computational details will be given in our journal version), and the performance of fault locationing for medium transmission lines with different phase angle asynchronism $\Delta\theta$. As shown in Fig. \ref{fig:sim_fm_bc}, prior to the fault occurrence, the values of $B$ and $C$ (especially $C$) increase as $\Delta\theta$ increases. When $\Delta\theta$ is equal to $25$, the value of $C$ before the fault occurrence is larger than $C$ after fault occurrence when there is no phase angle asynchronism. Therefore, the false alarm probability would be increased under TSA. As shown in Fig. \ref{fig:sim_fm_error}, the fault locationing error is proportional to $\Delta\theta$. When fault location index $D$ is equal to $0.5$ or $0.75$, the fault locationing error is as large as $0.3$ when $\Delta\theta$ is equal to $30$.

\subsubsection{Long Transmission Line}

\begin{figure}
  \centering
  \includegraphics[scale=0.35]{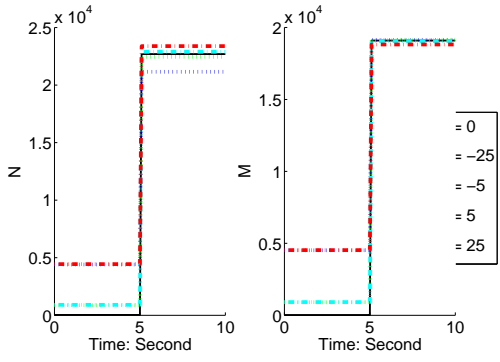}
  \caption{Fault indicators: N and M,  for long transmission lines}\label{fig:sim_fl_NM}
\end{figure}

\begin{figure}
  \centering
  \includegraphics[scale=0.35]{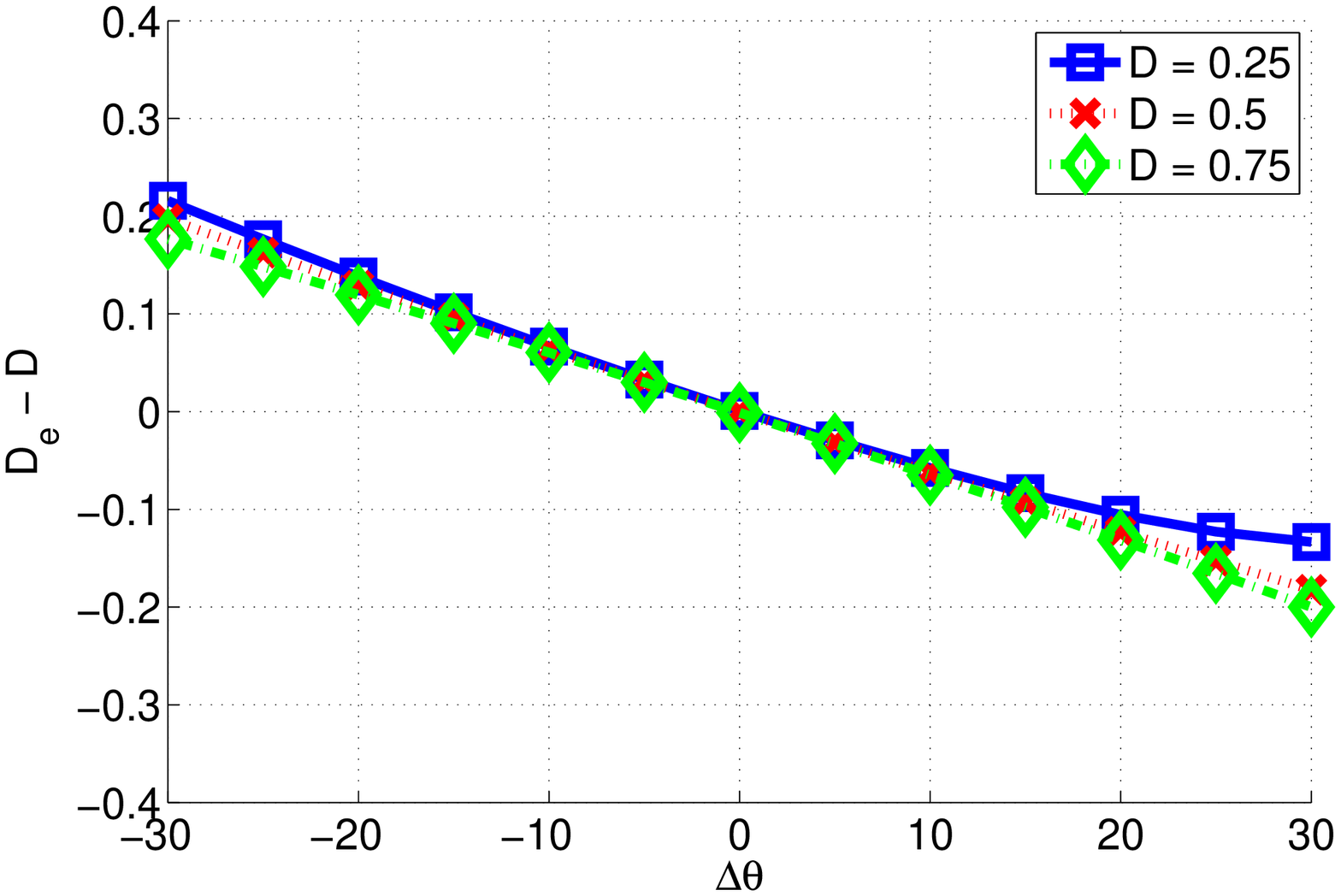}
  \caption{Performance of fault location for long transmission lines}\label{fig:sim_fl_error}
\end{figure}

\begin{figure}
  \centering
  \includegraphics[scale=0.35]{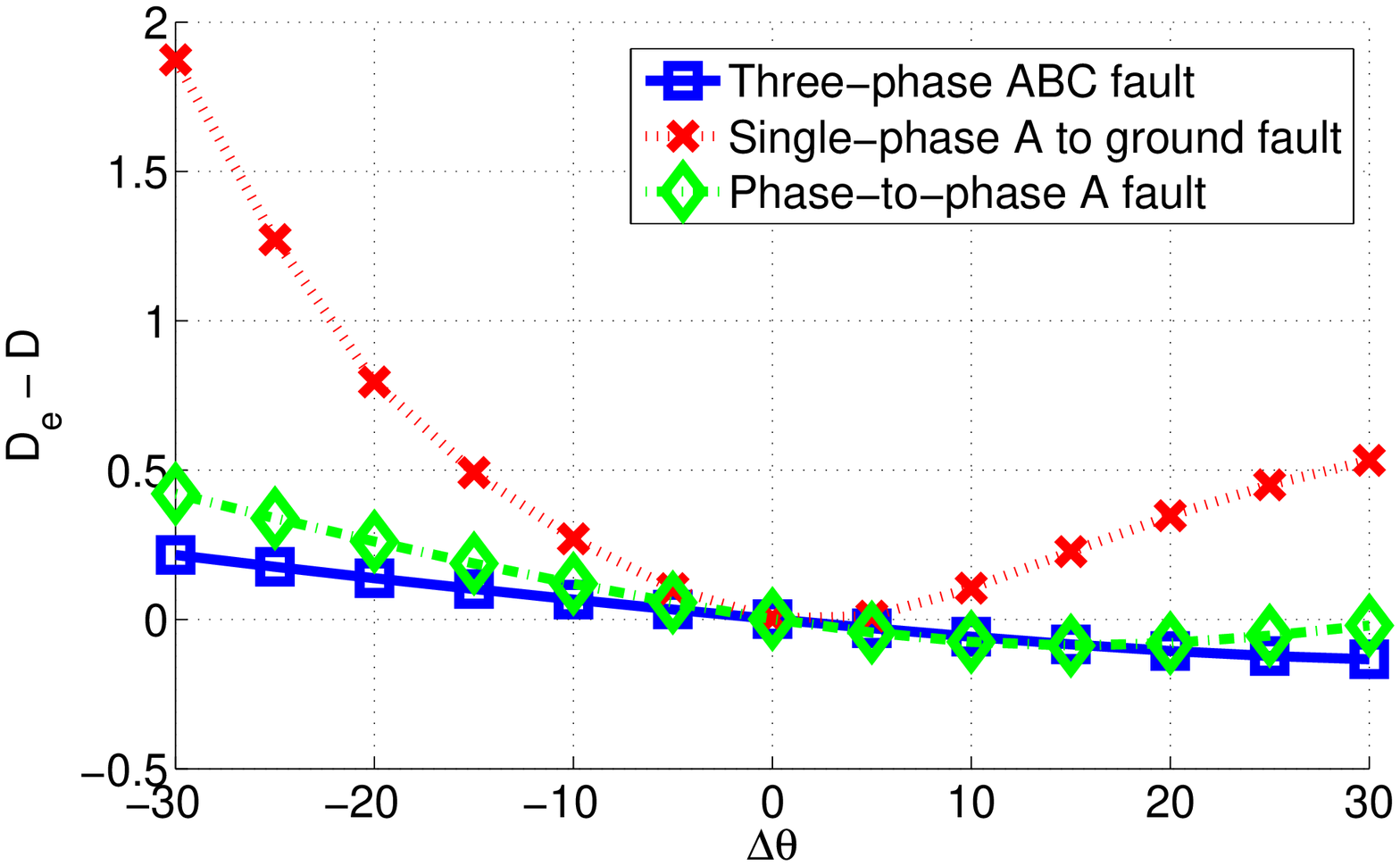}
  \caption{Performance of fault locationing for long transmissions line}\label{fig:sim_fl_error_dp}
\end{figure}

Fig. \ref{fig:sim_fl_NM} and Fig. \ref{fig:sim_fl_error} illustrate the fault indicators, $N$ and $M$, obtained from (\ref{eq:comp_M_N}) and the performance of fault locationing for long transmission lines with Phase ABC fault and different phase angle asynchronism $\Delta\theta$. Under TSA, the gaps of fault indicators, $N$ and $M$, decrease as $\Delta\theta$ increases. As the values of the fault indicators, $N$ and $M$, are much more than the fault indicators for short and medium transmission lines when fault occurs, the impact of TSA does not have much impact on the fault detection in long transmission lines. For long transmission lines, the fault locationing error is also proportional to the phase angle asynchronism $\Delta\theta$. When the fault location index $D$ is equal to $0.5$ or $0.75$, the fault locationing error is as large as $0.2$ when $\Delta\theta$ is equal to $30$. Fig. \ref{fig:sim_fl_error_dp} compares the performance of fault locationing with different types of faults under TSA. Fig. \ref{fig:sim_fl_error_dp} shows that, for type Phase A and type Phase AB faults, the performance of fault locationing is worse than that of type Phase ABC fault.

\subsection{Voltage Stability Monitoring}

\begin{figure}[htpb]
  \centering
  \includegraphics[scale=0.65]{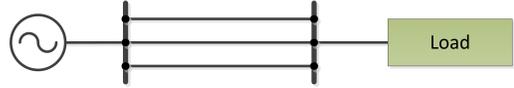}
  \caption{Simulation model for voltage stability}\label{fig:sim_v_model}
\end{figure}

The simulation model for the voltage stability monitoring is shown in Fig. \ref{fig:sim_v_model}. The root mean square amplitude of source voltage dynamically changes with frequency 1Hz. The load has a constant power. There are three transmission lines. A type phase ABC short-circuit fault occurs on transmission line 1 between 2s and 2.5s. Transmission lines 1 and 2 are tripped at time 4s and 6s, respectively.

\begin{figure}[htpb]
  \centering
  \includegraphics[scale=0.35]{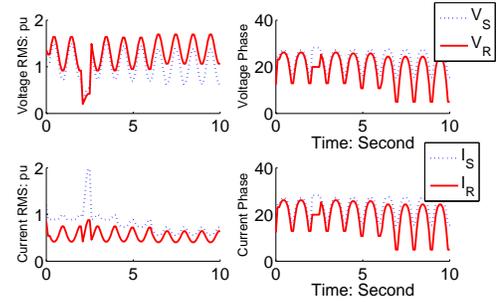}
  \caption{Voltages and currents at the sending and receiving ends}\label{fig:sim_vol_VI}
\end{figure}

\begin{figure}[htpb]
  \centering
  \includegraphics[scale=0.35]{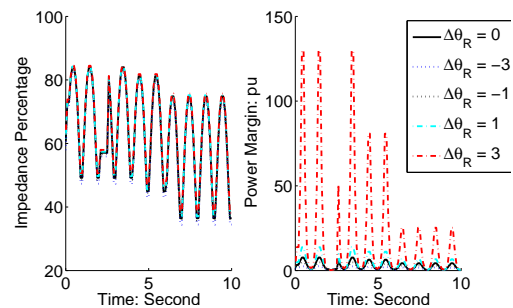}
  \caption{Voltage stability indices}\label{fig:sim_vol_MPz}
\end{figure}

\begin{figure}[htpb]
  \centering
  \includegraphics[scale=0.35]{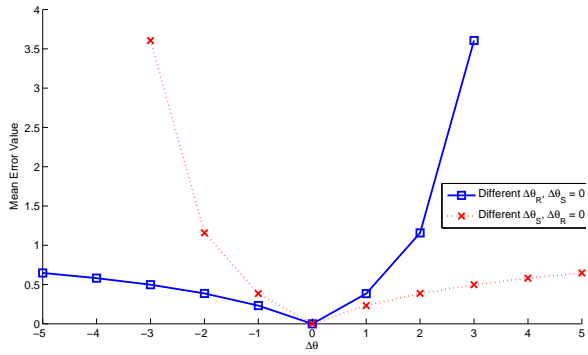}
  \caption{Performance of voltage stability monitoring index 1 with different phase angle asynchronism}\label{fig:sim_vol_Mp_d}
\end{figure}

The simulation results are shown in Figures \ref{fig:sim_vol_VI}, \ref{fig:sim_vol_MPz} and \ref{fig:sim_vol_Mp_d}, respectively. As shown in Fig. \ref{fig:sim_vol_MPz}, only the power margin index $\text{MARGIN}_P$ is affected by the phase angle asynchronism $\Delta\theta_R$ caused by TSA. Fig. \ref{fig:sim_vol_Mp_d} illustrates the normalized mean power margin index which is defined as
\begin{eqnarray}
E\left[|\widehat{\text{MARGIN}_P} - \text{MARGIN}_P|\right],
\end{eqnarray}
where $\widehat{\text{MARGIN}_P}$ is the estimated power margin index. As shown in Fig. \ref{fig:sim_vol_Mp_d}, the estimated error increases as $|\Delta\theta|$ increases. Another observation from the simulation result is that the estimated error is not symmetric with the phase angle asynchronism $\Delta\theta$. The increasing rate of estimated error for a positive $\Delta\theta$ is much larger than that for a negative $\Delta\theta$.

\subsection{Regional Disturbing Event Location}
\begin{figure}[htcp]
  \centering
  \includegraphics[scale=0.3]{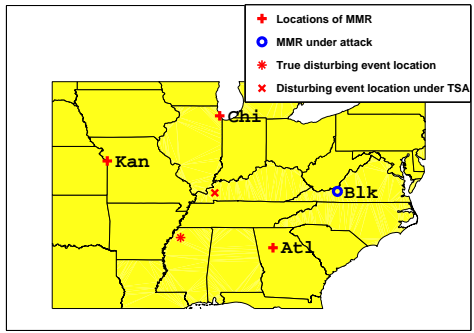}
  \caption{Simulation of TSA on disturbing event location}\label{Fig:Event_map}
\end{figure}

For the disturbing event location, the sampling is trigged by the GPS time signal as illustrated in Fig. \ref{Fig:SynMeasure}. A forged GPS time signal can control the sampling in a wrong time or provide a wrong time stamp for the measurements. The simulation on the effect on the event location is shown in Fig. \ref{Fig:Event_map}. It can be observed that, with one MMR under TSA, the estimation of disturbing event will be far away from the true position (the event happening in Mississippi is misled to Tennessee).

\section{Conclusion and Future Works}\label{sec: conclusion}
In this paper, we have identified the GPS spoofing based TSA in power grids. The time stamps are modified by the forged GPS signal, and the time stamp related measurements will be corrupted by TSA. TSA in several scenarios have been studied in this paper. For the transmission line fault detection and locationing, TSA can not only deteriorate the performance of fault locationing, but also increase the false alarm probability with some fault indicators. For the voltage stability monitoring, TSA can exaggerate the power margin and result in delaying or disabling the voltage instability alarm. It has also been demonstrated that the TSA can significantly damage the event location in power grid.

In our future work, we will study the protection scheme against TSA. From the viewpoint of signal processing, the fake GPS signal cannot erase the true GPS signal as illustrated in Fig. \ref{Fig:CorrPeak}. To mislead the GPS signal tracking, the spoofer must transmit a fake GPS signal with a higher SNR; thus we can detect the TSA by the SNR of the correlation peak. Since the spoofer's fake GPS signal has a significant direction of arrival, TSA may also be detected by applying the direction of arrival (DOA) discrimination \cite{Spoof_defense}, which will be further studied in our future work.

\section*{Acknowledgments}
This work was supported by the National Science Foundation
under grants CCF-0830451, ECCS-0901425 and CNS-1116826.


\end{document}